\documentclass{elsart}
\usepackage{epsfig}
\usepackage{amssymb}
\usepackage{amsmath}
\usepackage{subfigure}
\usepackage{float}
\usepackage{color}
\usepackage{hyperref}

\begin{document}
\begin{frontmatter}

\title{Vacuum Predictions and Measurements for an Internal Pellet Target}

\author[ISV]{I.~Lehmann\corauthref{cor}\thanksref{label1}}\ead{i.lehmann@physics.gla.ac.uk},
\author[ISV]{{\"O}.~Nordhage},
\thanks[label1]{Current address: Department of Physics \& Astronomy,
University of Glasgow,\\  Glasgow, G12 8QQ, Scotland, UK.}
\author[TSL]{C.-J.~Frid\'en}, 
\author[TSL]{G.~Norman},
\author[TSL]{C.~Ekstr\"om},
\author[ISV]{T.~Johansson},
 and
\author[ISV]{U.~Wiedner}

\corauth[cor]{Corresponding author.}
\address[ISV]{Department of Nuclear and Particle Physics, Uppsala University, 
  Box 535, SE-751~21 Uppsala, Sweden}
\address[TSL]{The Svedberg Laboratory, Uppsala University, Box 533, 
  SE-751~21 Uppsala, Sweden}

\begin{abstract}
Measurements with low Z targets at internal experiments typically
imply a gas load which deteriorates the ring vacuum.  Future
experiments need reliable estimates for the expected vacuum conditions
in order to design $4\pi$ detectors closely surrounding the
interaction area.

We present a method for the calculation of the resulting vacuum of
such a complex system using a Pellet Target.  In order to test the
method, a vacuum system with diagnostic tools has been set up and a
Pellet Target was operated under realistic conditions.  The results for 
the absolute vacuum agree within factors of two with the expected
pressures.
\end{abstract}

\begin{keyword}
Internal Target \sep Pellet Target \sep Evaporation 
\sep Vacuum Measurement \sep Vacuum Prediction \\

\PACS 29.25.Pj \sep 29.20.Dh \sep 68.03.Fg \\
\end{keyword}

\end{frontmatter}

\section{Introduction}\label{intro}

Modern storage rings need a high vacuum throughout the whole ring
to maintain the requirements on beam precision and lifetime.
Internal experiments covering $4\pi$ acceptance have typically little
space for vacuum pumps close to the interaction point. This is,
however, the region where a high gas load is introduced if windowless
Gas, Cluster, or Pellet Targets are used.  When planning a new
experimental facility the dimensions of the vacuum system have to be
adjusted such that an acceptable pressure is reached inside the
ring. In order to do this, it is crucial to have the means to predict the
vacuum taking the geometry and gas load by the target into account.

In this paper we present a method to calculate the vacuum in such a
system and compare it with measurements of the vacuum in a dedicated
test stand.  The measurements and calculations were done using a
vacuum system which is similar to the one anticipated for the future
PANDA experiment~\cite{TPR} and a Pellet Target as the source of gas.

A Pellet Target consists of micro-spheres (about 30$\,\mu$m in
diameter) of frozen gas called ``pellets''. Typically, hydrogen,
deuterium, or noble gases are used. The pellets travel with speeds of
50 to 100 metres per second and a spread of about 1\,mrad. Thus the
production can be metres away from the interaction point. Such a
target was used at the CELSIUS/WASA experiment and is currently in
operation at COSY~\cite{pellet,COSYWASA:04}.  A copy of the
generator of this target was set up as a testing facility at The
Svedberg Laboratory (TSL), Uppsala, Sweden.  The Pellet-Test Station
(PTS) has been equipped with a full vacuum system in order to perform
the measurements discussed below.

\section{Experimental Set-Up}\label{setup}

The Pellet-Test Station (PTS) is a fully operational target which has
been built to test and further develop the Pellet Target at
CELSIUS/WASA.  It is a completely independent system and the pellet
generator is largely a copy of the WASA target.  The details of the 
pellet generation can be found in Refs.~\cite{pellet}.
The basic principle is that gas, e.g.\ hydrogen, is liquefied in a
first stage and small droplets are formed by vibrations with 50 to
100\,kHz. These are then injected into vacuum which causes
freezing. The point of vacuum injection is often referred to as the
point of production, and we set our scale to it. To avoid large tails
of the distribution of the pellet stream and to define the width at
the interaction point, a skimmer is used at some point on the pellets'
path. Typically rates of several thousand pellets per second are
reached for a total width of the stream of about 2\,mm at an
interaction point about 2\,m below the vacuum injection.

\begin{figure}[htb]
 \begin{center}
 \includegraphics[width=0.6\columnwidth]{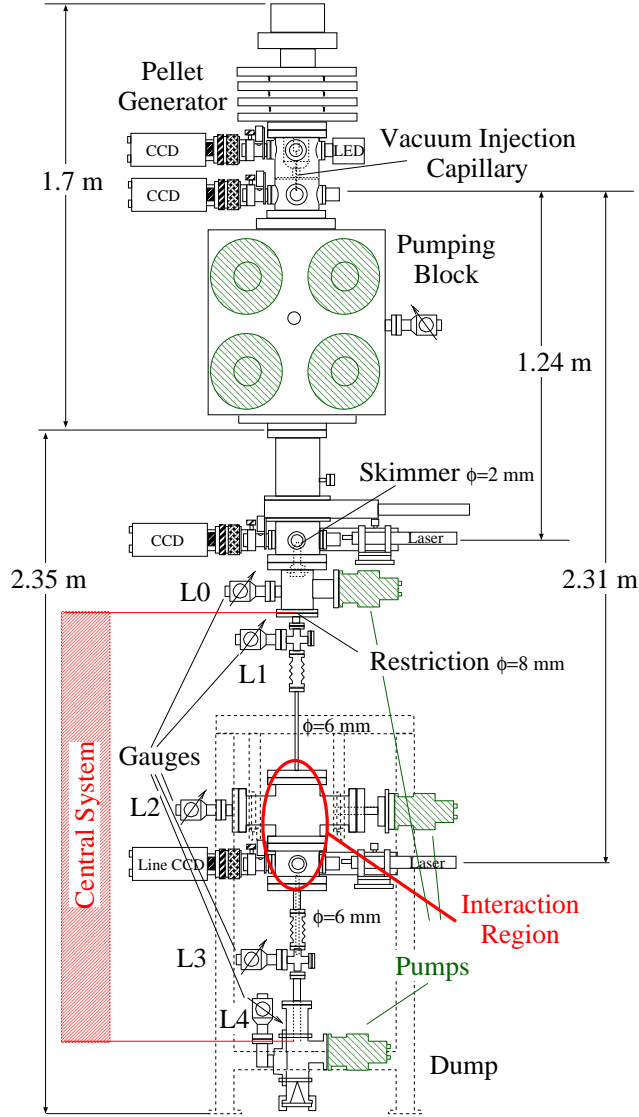}
 \caption{Sketch of the Pellet-Test Station (PTS). The Pellet
   Generator (top) ejects pellets at the Vacuum Injection Capillary,
   which pass a skimmer 1.24\,m below and reach the {\it Interaction
   Region} about one metre further below. Finally they are collected
   in the Dump (see also Table~\ref{t:PTS}).  The vacuum system
   resembles the situation inside a $4\pi$ detector, where the beam
   would enter the Interaction Region horizontally.}
 \label{fig:PTS}
 \end{center}
\end{figure}

The PTS differs by the following items from the original WASA
target~\cite{pellet}: 
\begin{itemize}
\item The cold head has been improved, leading to lower vibrations,
  faster pumping, better temperature and vacuum control.
\item The construction has been redesigned in order to allow for better
  access and faster exchange of parts on the generator, while keeping
  compatibility with the WASA system.
\item A new vacuum system has been added, which is designed to
  simulate the interaction region of PANDA vacuum-wise. It allows
  vacuum measurements at five points and visual observation of the pellets
  on two levels. Here cameras, counters and a line-scan camera have
  been installed~\cite{No06}.
\end{itemize}
A side view of the system is sketched in Fig.~\ref{fig:PTS} and
important parts and their distance to the production are listed in
Table~\ref{t:PTS}.  The pellet-generation system is mounted on top of
a block, which contains 4 turbo-molecular pumps with a pumping speed
of 2800\,l/s each.  It should be noted that all the pumping speeds
given in this document refer to the respective values for hydrogen, as
this is the gas we consider here. During operation the pressure in the
block is typically $10^{-5}$\,mbar.

The skimmer, a cone with a hole of 2\,mm that cuts the tails of the
pellet distribution, is positioned below the Pumping Block.  The upper
observation chamber holding this skimmer is used to align the pellet
stream such that a large fraction of the pellets pass through it.  A
side effect of such a skimmer is that some pellets are deflected only
slightly when hitting the rim or inner surface.  These continue with
the stream and may produce large gas loads, where the system is narrow
and difficult to pump.  In order to remove these, a second restriction
of 8\,mm diameter is placed 27\,cm below the skimmer. To pump the gas
away, the chamber between is equipped with a turbo-molecular pump with
150\,l/s and a vacuum gauge of Pirani type (L0).

When installing such a system at a large experimental set-up, the
skimmer would be placed much further up, e.g.\ in the middle of the
Pumping Block. In this case much more space would be available for
detectors while leaving the properties of the pellet stream
unchanged.\footnote{This is because the diameter of the skimmer can be
changed accordingly. Studies with several skimmer sizes have confirmed
the expected geometric relation without significant distortion.}

Below this point the components which simulate the vacuum system
inside a future detector follow. It is referred to as the {\it Central
System} in the following.  This corresponds also to the part which has
been modelled for the calculations discussed in Section~\ref{calc}.
There are two feed lines and an {\it Interaction Region}.  The feed
lines have a inner diameter of 6\,mm close to the interaction region
to simulate a typical experimental situation. Inside these lines,
vacuum gauges (L1 and L3) are installed to study the vacuum gradient
and systematics. At each of the three central measurement levels (L1,
L2, and L3) we have placed two vacuum gauges: one Pirani and one
Penning type. Thus a vacuum range from atmospheric pressure down to
well below $10^{-7}$\,mbar is covered. The gauges have been calibrated
and all values given in the following refer to hydrogen. The
uncertainty of any vacuum measurement is estimated to be 25\%.

The Interaction Region cannot be pumped effectively through the narrow
pellet-feed lines.  Moreover, assuming a moderate pumping speed at the
region itself, gas from the pellets is streaming into the Interaction
Region rather than being pumped away.  This is because the gas load
from the pellets is occurring all along their vertical path, i.e.\
mostly inside the feed pipes.  Both calculations and measurements
confirm that, indeed, the vacuum in the Interaction Region can be
improved using thinner rather than wider feed pipes.  Also a moderate
prolongation of these pipes would not lead to a significant change of
vacuum properties.

In an experimental set-up the beam pipe would intersect in the
Interaction Region.  Instead we accommodate two vacuum gauges (L2), a
turbo-molecular pump, and an observation house at this place.  The
pump (150\,l/s) is throttled by a conductance-limiting pipe such that
theoretically a pumping speed of 47\,l/s should be
reached.\footnote{In Section~\ref{calc} we will see that this value is
almost reached.}  Thus we expect to reproduce the pumping speed
foreseen at PANDA. The observation house is used to observe the
concentration of pellets. We use a $5$\,mm$ \times 0.1$\,mm laser and
two independent counters at this place to determine the count rate and
distribution of pellets~\cite{TPR,No06}. One counter uses light
reflected to angles of about 7$^\circ$, which is focused by a
lens-aperture system into a photo-multiplier.  A CCD line-scan camera
records light reflected to 90$^\circ$.  The rates recorded with both
systems agree to within 10\% and we use the first system for the
following analysis.

The Pellet Dump is placed at the bottom. A cone reflects pellets to
the sides to minimise the risk of a re-entry into the feed pipe. A
turbo-molecular pump with 210\,l/s pumping speed is mounted on one
side and a Pirani gauge (L4) monitors the vacuum. The system lacks an
active pellet-capturing mechanism (as e.g.\ active charcoal under
cryogenic conditions) and therefore backflow of vaporized hydrogen gas
is inevitable.  Furthermore, it cannot be excluded that some bouncing
pellets will find their way back upwards into the pipe system.  If
this happens, even a very small fraction will affect the vacuum at L3
significantly.

\begin{table}[tbh]
\begin{center}
\begin{tabular}{l c}
\hline \hline Component & Vertical position $[$cm$]$ \\
\hline
End of Vacuum-Injection Capillary           & 0           \\
Skimmer $\phi= 2\,$mm, observation windows  & 124         \\
Pirani gauge (L0), 150\,l/s pump            & 143         \\
Restriction: $\phi=8\,$mm                   & 151         \\
Pirani and Penning gauges (L1)              & 157         \\
Narrow pipe: $\phi=6\,$mm                   & 168 to 208  \\
Pirani and Penning gauges (L2), 150\,l/s pump & 213         \\
Observation windows                         & 231         \\
Narrow pipe: $\phi=6\,$mm                   & 238 to 251  \\
Pirani and Penning gauges (L3)              & 261         \\
Dump, Pirani gauge (L4), 210\,l/s pump      & 297 to 319  \\
\hline \hline
\end{tabular}
\end{center}
\caption{Major components of the PTS and their relative positions
  compared to the {\it Production Point} at the end of the
  Vacuum-Injection Capillary. The zone between 213 and 231\,cm is
  generally referred to as the {\it Interaction Region}. The {\it Central
  System} between 151 and 297\,cm is modelled by the vacuum
  calculations.}
\label{t:PTS}
\end{table}

\section{Vacuum Calculations}\label{calc}

To understand theoretically the vacuum in our system we use the
following input, which is described in greater detail below.
\begin{enumerate}
\item From the knowledge of the rate, size, and temperature of the
  pellets their gas load can be calculated~\cite{ON05}. \label{it:gas}
\item The conductances of the central system can be modeled using a
  one-dimensional approach based on Kirchoff-like rules for
  non-trivial pipe-systems in the regime of molecular
  flow. \label{it:cond}
\item We have experimentally obtained the pumping speeds at three
  positions of the system. \label{it:pump}
\end{enumerate}

In order to determine the gas load on the system from the outgassing
pellets, item~(\ref{it:gas}), we use the method in Ref.~\cite{ON05}
where a Helmholtz equation is solved in spherical coordinates.  First
we consider the mass loss per individual pellet and time
\begin{equation}
 \frac{d m_\mathrm{p}}{d t}
 = -1.668\, A\, \frac{P_\mathrm{S}-P_{\infty}}{\sqrt{2 \pi R_\mathrm{H} T}} 
 = (3.3 \pm 1.4)\, \mathrm{ng/s} \quad .
\end{equation}
Here, $P_\mathrm{S}$ is the solid-vapour saturation pressure from
Ref.~\cite{So86} and $R_\mathrm{H}$ is the specific gas constant of
hydrogen. We neglect the pressure of the surrounding vacuum
$P_{\infty}$ as it is much smaller than $P_\mathrm{S}$ in most cases.
With this assumption we tend to overestimate the gas load only at very
high rates.  The surface area $A$ is calculated from the pellet
diameter, which was determined to be on average $(45\pm3)\,\mu$m. This
is significantly larger than during typical pellet-target operation
due to the use of a 50\% larger nozzle than at WASA.  In
Ref.~\cite{ON05} the emissivity, which affects the equilibrium
temperature and mass loss, is an unknown parameter and was set to
$\epsilon = 0.5$.  In order to account for that uncertainty, we
attribute a 40\% error to the mass loss.

The total gas load to the system per second and per metre of travel is 
\begin{equation}
  \frac{d^2 G}{dt\,dy} = 
 - \frac{r_p}{v_p} \, \frac{d m_\mathrm{p}}{d t}
 \quad .
\end{equation}
where $v_p = 50 \pm 15\,$m/s is the experimentally determined average
pellet speed.  The errors from the emissivity and the speed of
the pellets dominate the total uncertainty of 50\%.  

To describe effectively the conductance, item~(\ref{it:cond}), we
limit ourselves to the relatively simple central system and consider
an effective pumping speed for that system for the final analysis.  A
one-dimensional model based on Kirchoff-like rules can be used to
describe a vacuum system with pipes of known conductance.  For a pipe
of circular cross section of diameter $d$ the conductance is
calculated using the method from Ref.~\cite{Cond}, in which a
characteristic conductance of an element is $ C_\mathrm{char} =
C_\mathrm{ap}a \, , $ where $ C_\mathrm{ap} = 0.347\, d^2 \,
\mathrm{l/(s\,mm^2)} $ refers to the aperture and for hydrogen gas at
298\,K. The transmission probability $a$ for a given pipe length $l$
has two extreme cases: in the limit when $l \ll d$ ({\it aperture}) $a$
equals 1, and for the other extreme case, when $l \gg d$ ({\it long
pipe}), the probability is $4d/3l$. Both limits, as well as
intermediate cases, are well described by the general-case probability
\begin{equation}
a_\mathrm{gen} =
\frac{14+4\frac{l}{d}}{14+18\frac{l}{d}+3(\frac{l}{d})^{2}} \quad .
\label{eq:a_gen}
\end{equation}

We use this approach and VAKLOOP~\cite{VAK} to model our central
system between the 8\,mm Restriction and the end of feed pipe in the
Dump, i.e.\ between 151\,cm and 297\,cm from the Production Point (see
shaded area in Figure~\ref{fig:PTS}).  For the calculation we assign
three effective pumping speeds to the system, one at each end ($S_1$
and $S_3$), and one at 213\,cm ($S_2$), where we have our throttled
pump simulating the interaction region of PANDA.

\begin{figure}[htb]
 \begin{center}
   \includegraphics[width=0.8\columnwidth]{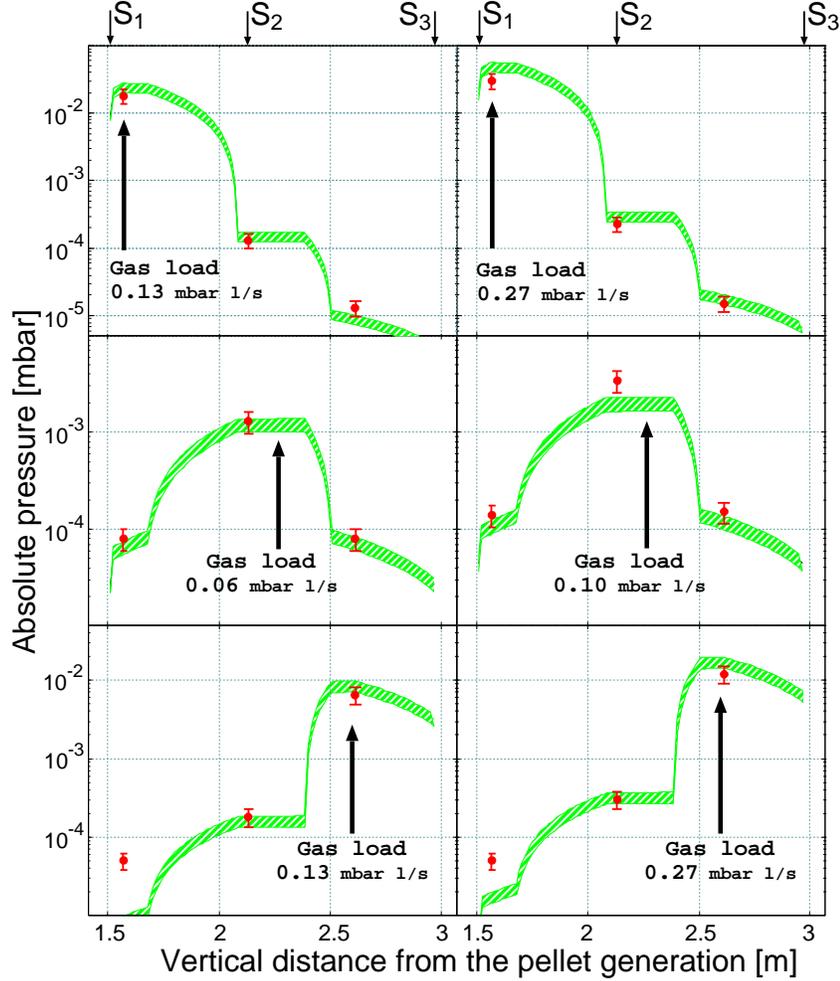}
  \caption{Measured vacuum (red points with error bars) when known
  flows of hydrogen gas are injected at three different locations
  indicated by the upward arrows in the panels.  The green shaded area
  shows an approach which only uses the three effective pumping
  speeds ($S_1, S_2$, and $S_3$) as free parameters (see text). The
  locations where those speeds are introduced are indicated at the top
  of the Figure.  The widths of the curves represent only the errors
  from the uncertainty of the gas input.  The errors of the points
  correspond to an assumed uncertainty of 25\% for the vacuum
  measurements.}
  \label{fig:Cal}
 \end{center}
\end{figure}

To obtain these effective pumping speeds, item~(\ref{it:pump}), we
have used the system itself and injected a known amount of gas.  A
thermal mass flow meter (V\"{o}gtlin, 2--100\,mln/min) calibrated for
hydrogen was used.  The error of 16\% given by the manufacturer is
shown as a band in Figure~\ref{fig:Cal}.  The procedure has been
performed for two gas loads and at three different positions (L1, L2,
and L3 levels).  Thus 18 experimental points for the vacuum have been
obtained, which were used in a single iterative approach with the
three pumping speeds as parameters (see Figure~\ref{fig:Cal}).

Though the absolute pressures vary by more than three orders of
magnitude, the values are described well within two $\sigma$ of the
errors.  The set of effective pumping speeds that reproduce the
experimental observations in Figure~\ref{fig:Cal} was found to be $S_1
= 10$\,l/s, $S_2 = 37$\,l/s, and $S_3 = 30$\,l/s.  From the geometry
alone it is difficult to calculate the pumping speeds for $S_1$ and
$S_3$, because the system is complicated and the pumps are
perpendicular to the openings.  Using a significantly simplified
geometry, values are obtained which are factors 5 and 3 higher than
obtained from the fit, respectively.  The $S_2$ value can be
calculated with much less ambiguity as the geometry is rather
simple. It is also the crucial value for the vacuum measurements
described in the following section. The resulting value of
$S_\mathrm{calc} = 47$\,l/s agrees within 20\% with the result of $S_2
= 37$\,l/s from the measurements.  This shows that the calculations
have to be restricted to geometrically rather simple arrangements.  We
use the measured $(S_1,S_2,S_3) = (10,37,30)\,$l/s as respective
pumping speeds at the three levels in the following.  The uncertainty
for these values is discussed in the next section.

\section{Results of the Vacuum Measurements}

\begin{figure}[htb]
 \begin{center}
 \includegraphics[width=0.8\columnwidth]{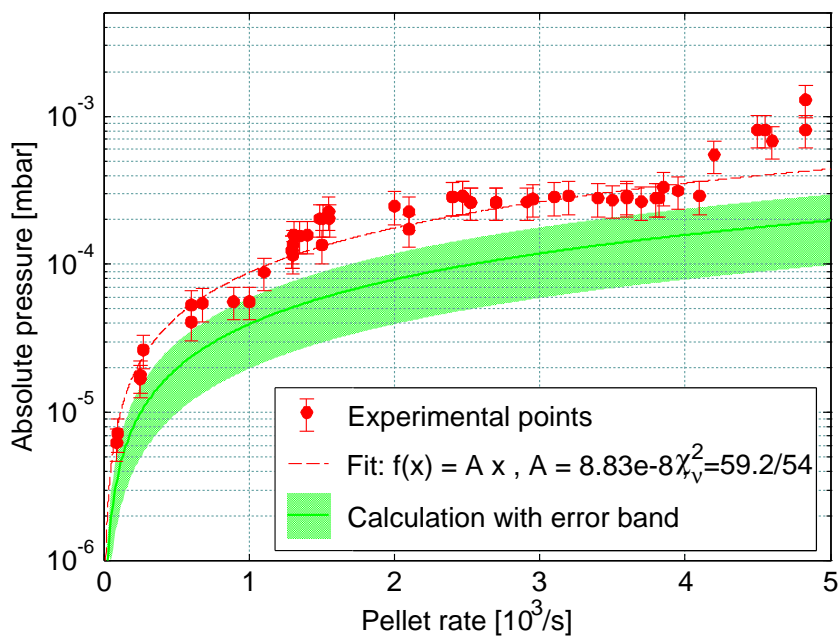}
  \caption{Measured vacuum as a function of pellet rate at the
  Interaction Chamber L2 (red points with error bars). The errors for
  the pellet rate are 20\% and not shown in order not to overcrowd the
  plot. The red thin dashed line is a linear fit to the data. The
  green line and shaded area show the vacuum expected from
  calculations and the error to this calculation, respectively.}
  \label{fig:Res}
 \end{center}
\end{figure}

The vacuum measurements were performed with the set-up described in
Section~\ref{setup}, where we assign a 25\% error to the vacuum
measurements. We used the pellet counter with the photo-multiplier
tube to measure the pellet rate. Here we have to assume an error of
about 20\%. We scanned a range from a few hundred to almost 5000
pellets per second while observing the pressure at all gauges.  The
pressure follows a linear relation at all levels, as expected (see
linear fit in Figure~\ref{fig:Res}).

At the Interaction Region (L2 gauge) the pressure increases to a few
times $10^{-4}$\,mbar when pellet rates of 4000/sec are reached (see
points with error bars in Figure~\ref{fig:Res}).  As already shown the
gas load from the pellets is proportional to their surface and thus
quadratically increasing with pellet size. As the pellets were 50\%
larger than at typical pellet target operation, we would expect the
vacuum to always stay below $10^{-4}$\,mbar with ``normal'' sized
pellets. Nevertheless, this clearly shows that the gas load is of
concern if only low pumping speeds, as foreseen at PANDA, can be
reached at the interaction point.

The solid and dotted lines in Figure~\ref{fig:Res} show the calculated
vacuum for the corresponding rate and their uncertainty, respectively.
The calculation has been done as described in Section~\ref{calc}
considering the errors for the mass loss and speed of the pellets.
Note that the uncertainty in pellet rate of 20\% is not shown
in the figure.  The calculations agree within two $\sigma$ but
systematically underestimate the pressures reached.  For L1 the values
agree within errors, while close to the dump at L3 the calculations
fail to describe the measured pressure.  These effects can be explained
by a contamination from bouncing pellets, that add to the gas load.
In the Interaction chamber this is less likely.  Thus we conclude that
systematic errors may be slightly larger than anticipated.  This may
be attributed to the uncertainties from the pumping speeds, which were
not taken into account.  A 30\% overestimation of the pumping speeds
determined by the method in Section~\ref{calc} would be sufficient to
reach full agreement within errors.

\section{Conclusions}

We have reported on the gas load that is expected when pellets are
used as an internal target.  We have presented a method to calculate
the vacuum when such a target is operated.  In order to check the
results of these calculations, measurements of the vacuum under
realistic conditions were performed at the Pellet-Test Station.  The
set-up was equipped with a dedicated system, which was calibrated in
terms of pumping speed.  The observed vacuum shows what has to be
expected at a future experiment with low pumping speed at the
interaction point.  The results of the calculations seem to lie
systematically too low, but are within factors of two in agreement with
the experiment.

In conclusion, a vacuum of about $10^{-4}$\,mbar has to be expected
when operating a pellet target with only 40\,l/s pumping speed at the
interaction point.  The model describes the vacuum resulting from the
operation of a pellet target in a complex system within factors of two.
This is especially remarkable as absolute values are calculated and
the measurements cover more than one order in magnitude in vacuum and
count rate.  Thus we expect that the vacuum for other configurations
may be predicted using that method within similar accuracy.  This may
be a very useful tool for the design of the vacuum systems of future
experiments, like PANDA.

\section*{Acknowledgements}
We would like to thank all the staff at The Svedberg Laboratory (TSL)
for their help with all kind of odds and ends, especially
Kjell~Fransson and the workshop personnel.  Acknowledged are also
Ingela~Nystr\"{o}m and Bo~Nordin (Centre of Image Analysis, Uppsala
University) for help with the image-analysis on the pellets. Also we
thank C.~Wilkin (University College, London) for reviewing the
document.  One of us ({\"O}.N.) also wish to thank the financial
support from GSI, Darmstadt, Germany.  This work was supported by the
European Community-Research Infrastructure Activity within the FP6
programme ``Structuring the European Research Area'' (Hadron Physics,
RII-CT-2004-506078).

\end{document}